\documentclass[sigconf]{acmart-me}

\usepackage{booktabs} 
\usepackage{url}
\usepackage{color}
\usepackage{enumitem}
\usepackage{balance}
\acmDOI{}

\acmISBN{}

\acmConference[MediaEval'20]{Multimedia Evaluation Workshop}{December 14-15 2020}{Online} 
\copyrightyear{}
\acmYear{}
\acmPrice{}

\acmPrice{}

\begin{document}
\title{Automatic Polyp Segmentation using U-Net-ResNet50}

\author{Saruar Alam\textsuperscript{1}, Nikhil Kumar Tomar\textsuperscript{2}, Aarati Thakur\textsuperscript{3},   Debesh Jha\textsuperscript{2,4}, Ashish Rauniyar\textsuperscript{5,6}}
\affiliation{\textsuperscript{1}University of Bergen, Norway\\ 
\textsuperscript{2}SimulaMet, Norway \\
\textsuperscript{3}Nepal Medical College, Kathmandu University, Nepal\\
\textsuperscript{4}UiT The Arctic University of Norway\\
\textsuperscript{5}University of Oslo, Norway \\
\textsuperscript{6}Oslo Metropolitan University, Norway \\
}

\email{saruar.alam@uib.no}

%
%
%
%
%

\renewcommand{\shorttitle}{MediaEval'20: Multimedia Evaluation Workshop}
\renewcommand{\shortauthors}{S. Alam et. al.}

\begin{abstract}
Polyps are the predecessors to colorectal cancer which is considered as one of the leading causes of cancer-related deaths worldwide. Colonoscopy is the standard procedure for the identification, localization, and removal of colorectal polyps. Due to variability in shape, size, and surrounding tissue similarity, colorectal polyps are often missed by the clinicians during colonoscopy. With the use of an automatic, accurate, and fast polyp segmentation method during the colonoscopy, many colorectal polyps can be easily detected and removed. The ``Medico automatic polyp segmentation challenge'' provides an opportunity to study polyp segmentation and build an efficient and accurate segmentation algorithm. We use the U-Net with pre-trained ResNet50 as the encoder for the polyp segmentation. The model is trained on Kvasir-SEG dataset provided for the challenge and tested on the organizer's dataset and achieves a dice coefficient of 0.8154, Jaccard of 0.7396, recall of 0.8533, precision of 0.8532, accuracy of 0.9506, and F2 score of 0.8272, demonstrating the generalization ability of our model.
\end{abstract}

%
%
%
%
%


\maketitle

\section{Introduction}
\label{sec:intro}
Identification and removal of polyps during colonoscopy have become a standard procedure. It is often challenging to detect polyps, as they are often hard to differentiate from surrounding normal tissue. These polyps are usually covered with stool, mucosa, and other materials that can obscure the correct diagnosis. This is especially true for the small, flat, and sessile polyps that are typically not visible during colonoscopy. Moreover, this increases the miss-rate of polyps up-to 25\%~\cite{kumar2017adenoma} and increases the risk of colorectal cancer in the affected patient. An increase in the 1\% adenoma detection rate leads to a 3\% decrease in the risk of colorectal cancer~\cite{corley2014adenoma}. Recently, deep learning techniques have been developed to overcome these challenges and improve polyp detection accuracy during colonoscopy. Polyp segmentation based deep learning methods has been successfully applied for automatic polyp detection in a real-time.

The automatic polyp segmentation plays an important role in the identification and localization of the polyps in the affected regions. It helps in analyzing the images or even video frames and classify each pixel into polyp or non-polyp class instances. This allows the clinician in easy, fast, and more accurate identification of the polyp in the affected region. The automated polyp segmentation can help in the development of a Computer-Aided Diagnosis (CADx) system, which is specially designed for colonoscopy procedures. 

The ``Medico Automatic Polyp Segmentation Challenge''~\cite{jha2020medico} consists of two tasks. The first task is ``Polyp segmentation task'' and the second is ``Algorithm efficiency task''.  We have submitted our model in task 1 only.

\begin{figure*}[h]
 \center
  \includegraphics[height=4cm]{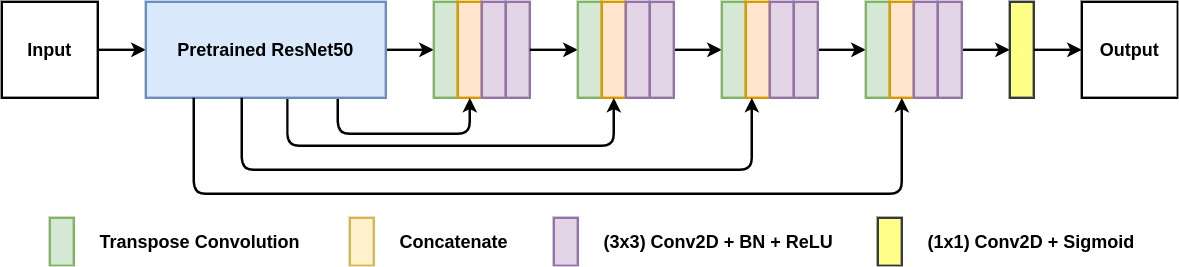}
  \caption{The proposed U-Net-ResNet50 architecture}
  \label{fig:architecture_fig}
\end{figure*}
\section{Related Works}
\label{sec:work}
For semantic segmentation task, encoder-decoder networks like FCN \cite{long2015fully}, U-Net \cite{ronneberger2015u}, etc are mostly preferred over other approaches. U-Net and its variants are used for both natural image segmentation and biomedical image segmentation. In general, the encoder uses multiple convolutions to learn and capture the essential semantic features ranging from low-level to high-level. These upscaled features are then concatenated with the features from the encoder using the skip connections and then followed by convolution layers to generate the final output in the form of a binary mask.

The encoder acts as a feature extractor,  where the decoder uses features extracted from the input to produce to desired segmentation mask. The encoder can be replaced by a pre-trained network such as VGG16 \cite{simonyan2014very}, VGG19 \cite{simonyan2014very},  etc. These pre-trained networks are already trained on the ImageNet \cite{russakovsky2015imagenet} dataset and have the necessary feature extraction capabilities. Architectures like SegNet \cite{badrinarayanan2017segnet} and TernausNet \cite{iglovikov2018ternausnet} use pre-trained VGG16 and VGG11 respectively for segmentation task. 

With the success of the residual network \cite{he2016deep}, ResNet50 is one of the commonly used architecture for any transfer learning task. The residual network uses two $3 \times 3$ convolutional layers and an identity mapping. Each convolution layer is followed by a batch normalization layer and a Rectified Linear Unit (ReLU) activation function. The identity mapping is the shortcut connection connecting the input and output of the convolutional layer. The identity mapping helps in building a deeper neural network by eliminating the problem of vanishing gradients and exploding gradients.

\section{Approach}
\label{sec:approach}
Figure~\ref{fig:architecture_fig} shows an overview of the proposed U-Net-ResNet50 architecture. It is an encoder-decoder based architecture, where ResNet50 trained on ImageNet dataset \cite{russakovsky2015imagenet} is used . The use of a pre-trained encoder helps the model to converge easily.
The input image is fed into the pre-trained ResNet50 encoder, consisting of a series of residual blocks as their main component. These residual blocks help the encoder extract the important features from the input image, which are then passed to the decoder. The decoder starts a transpose convolution that upscales the incoming feature maps into the desired shape. Next, these upscaled feature maps are concatenated with the specific shape feature maps from the pre-trained encoder via skip connections. These skip connections help the model to get all the low-level semantic information from the encoder, which allows the decoder to generate the desired feature maps. After that, it is followed by the two $3 \times 3$ convolution layer, where each layer is followed by a batch normalization layer and a ReLU non-linearity. The last decoder block's output is passed to a $1 \times 1$ convolution layer, which is further passed to a sigmoid activation function, finally generating the desired binary mask.\\
\indent The FastAI (version 2.0) library \cite{docsfast} is used to train and evaluate our model. We have employed resizing, flipping, rotating, zooming, lightning, warping, and normalizing intensity based on the ImageNet dataset to augment the input images for training. The model uses Adam optimizer with an initial learning rate of $10^{-2}$, and cross-entropy loss as its loss function. We have employed the one-cycle policy where the learning rate changes during training and achieves super-convergence \cite{smith2019super}. We have run just 50 epochs for training, and the model has converged.

\section{Results and analysis}
The Medico Automatic Polyp Segmentation challenge~\cite{jha2020medico} provides an opportunity to study the potential and challenges of automated polyp segmentation. This study aims at building a model that performs well on the organizer's dataset while training on a separate Kvasir-SEG dataset \cite{jha2020kvasir}.\\
\indent Table~\ref{table:quantative_results} shows the overall results of the U-Net-ResNet50 architecture on the Kvasir-SEG test dataset and the organizer's test dataset provided for the final evaluation of the model. For the evaluation of the model, the Jaccard index, S{\o}rensen-Dice coefficient (DSC), recall, precision (Prec.), accuracy (Acc.), and the F2 are used as the evaluation metrics. Our trained U-Net-ResNet50 model achieved a dice coefficient of 0.8154, Jaccard of 0.7396, recall of 0.8533, precision of 0.8532, accuracy of 0.9506 and F2 score of 0.827 on the organiser's test dataset  which can be seen from the table ~\ref{table:quantative_results}. These results demonstrate the generalization ability of our model. Moreover Table~\ref{table:quantative_results} also shows that the recall value of the organizer's test dataset is 1.00\% higher than the Kvasir-SEG test dataset. This shows that the model is not overfitting. 


\begin{table}[t]
 \caption{Quantitative Results on Kvasir-SEG and Test Set (Challenge) Dataset for Task 1.}
    \label{table:quantative_results}
    \def\arraystretch{2.0}
     \setlength\tabcolsep{2.2pt}
    \centering
  \begin{tabular}{@{}l|l|l|l|l|l|l@{}}
\hline
\textbf{Dataset}& \textbf{Jaccard} & \textbf{DSC} & \textbf{Recall} & \textbf{Prec.} & \textbf{Acc.} & \textbf{F2} \\ \hline
Kvasir-SEG & 0.7871 & 0.8926 & 0.8433 & 0.9207 & 0.9639 & 0.8585 \\ \hline
Test Set     & 0.7396 & 0.8154 & 0.8533 & 0.8532 & 0.9506 & 0.8272  \\ \hline
\end{tabular}
\end{table}


\section{Conclusion \& Future Work}
With our U-Net-ResNet50, we achieved competitive performance on the organizer's dataset with a dice coefficient of 0.8154. By replacing the U-Net encoder with a pre-trained ResNet50 and employing a one-cycle policy during training,  we are able to converge the model in a short time. Thus, it helps in reducing the training time as the encoder weights are not initialized from scratch. This is an important step towards faster convergence, which would be useful when the availability of high-performance computing resources is limited. \\
\indent In the future, we would like to experiment with more than one pre-trained encoder by fusing their feature maps and using them for training our model.  

\begin{acks}
The computations in this paper were performed on the equipment provided by the Experimental Infrastructure for Exploration of Exascale Computing (eX3), which is financially supported by the Research Council of Norway under the contract 270053.\\
\indent The authors would also like to thank the machine learning group of Mohn Medical Imaging and Visualization (MMIV) Centre, Norway, for providing the computing infrastructure for the experiments. 
\end{acks}

\newpage
\bibliographystyle{ACM-Reference-Format}
\def\bibfont{\small} 
\balance
\bibliography{sigproc} 

\end{document}